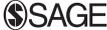

# Nonlinear dynamics of a functionally graded piezoelectric micro-resonator in the vicinity of the primary resonance

Meysam T Chorsi[1], Saber Azizi[2] and Firooz Bakhtiari-Nejad[3]

## Abstract
This research is on the nonlinear dynamics of a two-sided electrostatically actuated capacitive micro-beam. The micro-resonator is composed of silicon and PZT as a piezoelectric material. PZT is functionally distributed along the height of the micro-beam according to the power law distribution. The micro-resonator is simultaneously subjected to DC piezoelectric and two-sided electrostatic actuations. The DC piezoelectric actuation leads to the generation of an axial force along the length of the micro-beam and this is used as a tuning tool to shift the primary resonance of the micro-resonator. The governing equation of the motion is derived by the minimization of the Hamiltonian and generalized to the viscously damped systems. The periodic solutions in the vicinity of the primary resonance are detected by means of the shooting method and their stability is investigated by determining the so-called Floquet exponents of the perturbed motions. The basins of attraction corresponding to three individual periodic orbits are determined. The results depict that the higher the amplitude of the periodic orbit, the smaller is the area of the attractor.

## 1. Introduction

Today the application of microelectromechanical systems (MEMS) has increased considerably and therefore the study of their behavior is of great importance. The initial papers that were published on MEMS account for the MEMS statics; however recently reported reviews/papers are mostly on the dynamics of MEMS. A group of papers in the literature on the dynamics of capacitive MEM structures are devoted to the so-called pull-in instability due to a DC voltage (Osterberg, 1995; Rezazadeh et al., 2006; Azizi et al., 2011). Another field of research which has gained a great deal of attention is on the nonlinear dynamics of capacitive micro-structures subject to a combination of a bias DC and a harmonic excitation. This type of actuation enables the design of low driving voltage and switching time (Radio Frequency) RF switches (Nayfeh et al., 2007a). Younis and Nayfeh (2003) studied the nonlinear response of a resonant micro-beam subject to an electric actuation; using a single mode approximation they performed the perturbation method to capture the amplitudes of the periodic solutions and studied the stability of the periodic orbits. They also reported that a 1:3 internal resonance between the first two modes may also be activated when the first mode is excited in the vicinity of the primary resonance. Nayfeh et al. (2007a) studied the dynamics of a one-sided electrostatically actuated MEMS resonator actuated by a combination of a DC and harmonic voltage. They used the shooting method to capture the periodic orbits in the vicinity of the primary resonance and studied the stability of the periodic orbits by means of determining the eigenvalues of the so-called monodromy matrix; based on the amount of

[1]Mechanical Engineering Department, Amirkabir University of Technology, Tehran, Iran
[2]Urmia University of Technology, Urmia, Iran
[3]Amirkabir University of Technology, Tehran, Iran

**Corresponding author:**
Saber Azizi, Urmia University of Technology, Band Road, Urmia 57166-9318, Islamic Republic of Iran.
Email: s.azizi@mee.uut.ac.ir

the DC voltage, they reported both softening and hardening effects in the frequency response curves. Younis et al. (2003) studied the nonlinear dynamics of the same model proposed by Nayfeh et al. (2007a), and they proposed a novel approach to generate reduced-order models; they reported that a Taylor series expansion of the nonlinear electrostatic term fails to represent the electrostatic force especially in the vicinity of the pull-in instability. Abdel-Rahman et al. (2003b) studied the same model as proposed by Younis et al. (2003, 2007), Nayfeh et al. (2005a,b; 2007a), and Azizi et al. (2014). They simulated the dynamic behavior and compared the results with numerical solutions and available experimental results in the literature. Najar et al. (2010) studied both the primary and secondary resonances of the first mode; they used a discretization technique to combine the differential quadrature method (DQM) and the finite difference method for the space and time respectively. They accounted for two main sources of nonlinearities including geometrical and displacement-dependent electro-static force. They reported pull-in bands, bifurcation types, and stability types in the frequency response curves. Abdel-Rahman et al. (2003a), and Abdel-Rahman and Nayfeh (2003) published papers on the sub and super harmonic resonances of a clamped-clamped capacitive micro-beam subject to the same actuation voltage as reported in Abdel-Rahman and Nayfeh (2003), Younis et al. (2003), Nayfeh et al. (2005a,b; 2007a), Nayfeh and Younis, 2005, and Azizi et al. (2014). They showed that based on the amount of applied DC voltage and quality factor, the micro-resonator may have a region of multi-valued solutions and two bifurcation points which leads to hysteretic behavior. In another paper Nayfeh et al. (2005a) investigated the hardening/softening behavior in the vicinity of primary resonance of a fully clamped capacitive micro-beam. Nayfeh and Younis (2005) represented analysis of a capacitive clamped-clamped micro-beam under secondary resonance excitation. They reported that because the frequency response curves have very steep passband-to-stopband transitions, this makes the combination of a DC voltage and a sub-harmonic excitation in the order of one-half a very good candidate for designing high-sensitive RF MEMS filters. Alsaleem et al. (2009) presented analysis, and experimental investigation for non-linear resonances and the dynamic pull-in instability in electrostatically actuated resonators. They determined the basins of attraction for the periodic attractors on the frequency response curves. Lin and Zhao (2003, 2005a,b) determined the influence of Casimir force on the nonlinear behavior of an electrostatically actuated nano-switch. They determined the bifurcation properties of the nano-switch and also the periodic orbits on the phase plane.

Up to now, many researchers have focused on the dynamics of capacitive clamped-clamped micro-beams; two main sources of nonlinearities including mid-plane stretching, known as geometric nonlinearity and displacement dependent electrostatic force, govern the dynamics of fully clamped capacitive micro-beams. The governing nonlinear terms add interesting behaviors such as multi-value response in the vicinity of primary, secondary and internal resonances (Abdel-Rahman and Nayfeh, 2003; Abdel-Rahman et al., 2003a,b; Younis and Nayfeh, 2003; Nayfeh et al., 2007a; Azizi et al., 2014), and chaotic behavior (Azizi et al., 2013) to the dynamics of these structures. Azizi et al. (2014) proposed a fully-clamped capacitive micro-beam subject to a combination of DC-AC actuation; the micro-beam was silicon based sandwiched with piezoelectric layers. With piezoelectric actuation not only could they tune the primary resonance of the micro-resonator but also they could cancel the pull-in band on the frequency response curves. Although piezoelectric actuation leads to the tuning of the operating frequency of the resonator, the stress concentration and detachment of the piezoelectric layers due to rapid change of the mechanical properties is a challenging dilemma; therefore application of functionally graded piezoelectric (FGP) MEMS as a continuum media (Azizi et al., 2012) is an alternative to take advantage of the tunability of the device as well as overcoming the piezoelectric deposition limitations.

In the present study a two-sided electrostatically actuated fully-clamped FGP/silicon micro-beam resonator is investigated. The dynamics of micro beams subject to two-sided electrostatic actuation is more complicated than the single electrostatic actuation due to the appearance of chaotic dynamics which is focused on especially in micro mixers (Suzuki and Ho, 2002; Haghighi and Markazi, 2010; Azizi et al., 2013). Furthermore the pull-in threshold due to bias DC voltage occurs in higher voltages compared with the single electrostatic actuation (Azizi et al., 2011, 2013). The tunability of the pioneer resonant sensors is due to the DC voltage of single electrostatic actuation, however this makes the device tunable only in the backward direction. The proposed model is a tunable (both in forward and backward directions) resonant sensor. The novelty of the model is its tunability of resonance frequency and its robustness against pull-in instability due to the two-sided simultaneous electrostatic actuation. Furthermore the piezoelectric deposition limitations on the pure silicon micro-beam have been cancelled. The study accounts for the dynamics of the micro-resonator in the primary resonance condition. The frequency response curves for various portions of silicon/piezoelectric are determined. By means of application of an appropriate piezoelectric voltage, the operating resonance frequency is tuned. The periodic orbits are captured using the so-called shooting technique and their stability is investigated by determining the eigenvalues of the monodromy matrix. The basins of attraction for some periodic orbits are determined on the phase plane.

## 2. Modeling

The proposed resonator is an FGP fully clamped capacitive micro-beam of length $l$, thickness $h$, and width $a$. The micro-beam is composed of silicon and PZT as a piezoelectric material and the mechanical properties are distributed according to power law distribution along the thickness of the micro-beam. The resonator is subject to two stationary electrodes located on either side of the micro-beam. Through the upper electrode a combination of DC and a harmonic voltage with amplitude $V_{AC}$ and frequency $\Omega$ is applied. The lower electrode imposes a pure DC voltage the same as that of the upper electrode (Haghighi and Markazi, 2010; Azizi et al., 2013). As depicted in Figure 1(a) DC voltage is applied to the upper and lower layers of the FGP micro-beam which leads to the introduction of an axial force along the length of the micro-beam; this is used as a tool to tune the primary resonance frequency in this study. To create a uniform electric field, a pure conductive layer in the upper and lower planes of the micro-beam is required; to accomplish this, a thin film metallic layer is deposited to the corresponding surfaces without considerably affecting the governing differential equation.

The coordinate system is attached to the mid plane of the left clamped end of the micro-beam. For a definite $z$, the mechanical properties are assumed to be a linear combination of the portion of silicon and that of piezoelectric material (Azizi et al., 2012). The mechanical properties associated with silicon and PZT are denoted by subscriptions '$S$' and '$P$' respectively. The mechanical properties vary with respect to power law distribution as:

$$E(z) = E_q e^{\gamma|z|}$$
$$\rho(z) = \rho_q e^{\alpha|z|} \quad (1)$$
$$e_{31}(z) = e_{(31_P)} e^{\mu|z|} - \beta$$

where $E_q, \gamma, \rho_q, \alpha, \mu$ and $\beta$ are constants which are determined based on the following conditions (Azizi et al., 2012):

$$z = 0: \quad MP = MP_0 = P_{S_0} MP_S + P_{P_0} MP_P$$
$$z = \frac{h}{2}: \quad MP = MP_u = P_{S_u} MP_S + P_{P_u} MP_P \quad (2)$$

where $MP$ stands for any functionally graded mechanical property including elasticity modulus, density or piezoelectric coefficient. $P_{S_0}$ and $P_{P_0}$ represent the proportion of silicon and piezoelectric material on the midplane ($z = 0$); accordingly $P_{S_u}$ and $P_{P_u}$ correspond to those of the upper and lower plane ($z = \pm h/2$).

Considering the conditions in equation (2), equation (1) reduces to (Azizi et al., 2012):

$$E(z) = (E_0) e^{\frac{2}{h}|z|\ln(\frac{E_u}{E_0})}$$
$$\rho(z) = (\rho_0) e^{\frac{2}{h}|z|\ln(\frac{\rho_u}{\rho_0})} \quad (3)$$
$$e_{31}(z) = e_{31_P}(e^{\frac{2}{h}|z|\ln(1-P_{P_0}+P_{P_u})} - 1 + P_{P_0})$$

The total potential strain energy of the micro-beam is due to bending ($U_b$), axial force due to piezoelectric actuation ($U_p$), axial force due to the mid-plane stretching ($U_a$), and electrical co-energy stored between the stationary electrodes and the micro-beam ($U_e^*$) (Azizi et al., 2014).

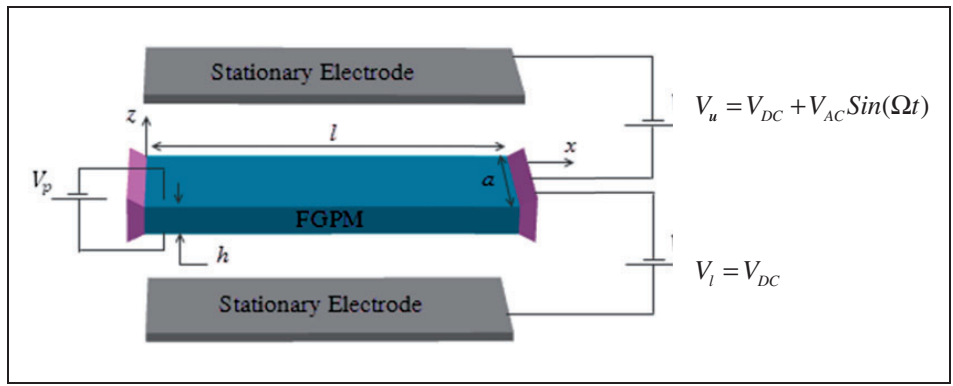

**Figure 1.** Schematics of the functionally graded piezoelectric micro-resonator (FGPM) and the applied voltages.

The strain energy due to mechanical bending is expressed as

$$U_b = \int \frac{\varepsilon_b \sigma_b}{2} dv$$
$$= \int \frac{E}{2}\left(-z \frac{\partial^2 w}{\partial x^2}\right)\left(-z \frac{\partial^2 w}{\partial x^2}\right) dv \quad (4)$$

where $\varepsilon_b$ and $\sigma_b$ are the strain and stress fields due to the bending and $dv$ is the volume of infinite small element; considering the geometry of the micro-beam and to simplify, equation (4), reduces to

$$U_b = \frac{(EI)_{eq}}{2} \int \left(\frac{\partial^2 w}{\partial x^2}\right)^2 dx \quad (5)$$

where (Azizi et al., 2012, 2014):

$$(EI_{yy})_{eq} = \int_{-\frac{h}{2}}^{\frac{h}{2}} aE(z)z^2 dz \quad (6)$$

The clamped-clamped boundary condition imposes the extended length of the beam ($l'$) to become more than the initial length $l$ which leads to the introduction of an axial stress and accordingly an axial force denoted as (Azizi et al., 2013)

$$F_a = \frac{(EA)_{eq}}{l}(l' - l) \approx \frac{(EA)_{eq}}{2l}\int_0^l \left(\frac{\partial w}{\partial x}\right)^2 dx \quad (7)$$

where (Azizi et al., 2012, 2014):

$$(EA)_{eq} = \int_{-\frac{h}{2}}^{\frac{h}{2}} aE(z) dz \quad (8)$$

Considering $L \gg w$, the stretched length ($l'$) is obtained by integrating the arc length "$ds$" as (Zhang and Zhao, (2003)):

$$l' = \int_0^l ds \approx \int_0^l \sqrt{1 + \left(\frac{\partial w}{\partial x}\right)^2} dx = l + \frac{1}{2}\int_0^l \left(\frac{\partial w}{\partial x}\right)^2 dx \quad (9)$$

The corresponding strain energy due to the mid-plane stretching is:

$$U_a = \frac{1}{2} F_a (l' - l) \quad (10)$$

Substituting equations (7) and (9) into equation (10), the strain energy due to the mid-plane stretching reduces to

$$U_a = \frac{(EA)_{eq}}{8l}\left(\int_0^l \left(\frac{\partial w}{\partial x}\right)^2\right)^2 \quad (11)$$

Immovable boundary conditions prompt the piezoelectric actuation to result in another axial force; based on the constitutive equation of piezoelectricity, (Azizi et al., 2012) and considering the direction of the applied electrical field ($E_3$) the axial stress due to the piezoelectric actuation reduces to (Azizi et al., 2014)

$$\sigma_1 = -e_{31} E_3 \quad (12)$$

where $e_{31}$ is the corresponding piezoelectric voltage constant (Coulomb/m$^2$); considering $E_3 = V_p/h_p$, the axial force due to the piezoelectric actuation reduces to (Azizi et al., 2014)

$$F_p = \int_{A_p} \sigma_1 dA_p = \frac{2aV_p}{h}\int_0^{h/2} -e_{31}(z) dz \quad (13)$$

The strain potential energy due to the axial piezoelectric force is

$$U_p = F_p(l' - l) = \frac{F_p}{2}\int_0^l \left(\frac{\partial w}{\partial x}\right)^2 dx \quad (14)$$

The total potential energy of the system is as follows:

$$U = U_b + U_a + U_p \quad (15)$$

The kinetic energy of the micro-beam is represented as

$$T = \frac{(\rho ah)_{eq}}{2} \int_{x=0}^{x=l} \left(\frac{\partial w}{\partial t}\right)^2 dx \quad (16)$$

where

$$(\rho ah)_{eq} = \int_{-\frac{h}{2}}^{\frac{h}{2}} a\rho(z) dz \quad (17)$$

The work of the electrostatic force from zero deflection to $w(x)$ is expressed as

$$w_{el} = \int_0^l \left(\int_0^w \frac{\varepsilon_0 a V_u^2}{2(g_0 - \zeta)^2} d\zeta\right) dx + \int_0^l \left(\int_0^w \frac{-\varepsilon_0 a V_l^2}{2(g_0 + \zeta)^2} d\zeta\right) dx$$
$$= \frac{\varepsilon_0 a V_u^2}{2}\int_0^l \left(\frac{1}{g_0 - w} - \frac{1}{g_0}\right) dx + \frac{\varepsilon_0 a V_l^2}{2}\int_0^l \left(\frac{1}{g_0 + w} - \frac{1}{g_0}\right) dx \quad (18)$$

where $g_0$ is the initial gap between the micro-beam and the substrate and $\varepsilon_0$ is the dielectric constant of the gap medium and $\zeta$ is a dummy parameter. The governing partial differential equation of the motion is obtained by the minimization of the Hamiltonian using variational principle as

$$\delta \int_0^t H dt = \delta \int_0^t (T - U + w_{el}) dt = 0 \quad (19)$$

Introducing equations (15), (16) and (18) into equation (19), the Hamiltonian reduces to:

$$\delta \int_0^t H dt$$

$$= \int_0^t \begin{Bmatrix} -(EI)_{eq} w'' \delta w' \big|_0^l + (EI)_{eq} w''' \delta w \big|_0^l \\ -(EI)_{eq} \int_0^l w^{IV} \delta w dx \\ -F_p w' \delta w \big|_0^l + F_P \int_0^l w'' \delta w dx \\ -\dfrac{(EA)_{eq}}{2} \int_0^l w'^2 dx w' \delta w \big|_0^l \\ +\dfrac{(EA)_{eq}}{2} \int_0^l w'^2 dx \int_0^l w'' \delta w dx \\ +\dfrac{\varepsilon_0 a V_u^2}{2} \int_0^l \dfrac{\delta w}{(g_0 - w)^2} dx \\ -\dfrac{\varepsilon_0 a V_l^2}{2} \int_0^l \dfrac{\delta w}{(g_0 + w)^2} dx \end{Bmatrix} dt$$

$$+ \int_0^t \left\{ (\rho A)_{eq} \int_0^l \dot{w} \delta w \big|_0^t - (\rho A)_{eq} \int_0^l \ddot{w} \delta w dt \right\} dx = 0 \quad (20)$$

The governing equation and the corresponding boundary conditions reduce to

$$(EI)_{eq} \dfrac{\partial^4 w(x,t)}{\partial x^4} + (\rho A)_{eq} \dfrac{\partial^2 w(x,t)}{\partial t^2}$$
$$- \left( F_P + \dfrac{(EA)_{eq}}{2l} \int_0^l \left( \dfrac{\partial w(x,t)}{\partial x} \right)^2 dx \right) \dfrac{\partial^2 w(x,t)}{\partial x^2} \quad (21)$$
$$= \dfrac{\varepsilon_0 a (V_{DC} + V_{AC} \sin(\Omega t))^2}{2(g_0 - w(x,t))^2} - \dfrac{\varepsilon_0 a V_{DC}^2}{2(g_0 + w(x,t))^2}$$

Subjected to the following boundary conditions:

$$w(0,t) = w(l,t) = 0, \quad \dfrac{\partial w(0,t)}{\partial x} = \dfrac{\partial w(l,t)}{\partial x} = 0 \quad (22)$$

To obtain the governing differential equation of the motion in its non-dimensional form the following non-dimensional parameters are used:

$$\hat{w} = \dfrac{w}{g_0}, \quad \hat{x} = \dfrac{x}{l}, \quad \hat{t} = \dfrac{t}{\tilde{t}}, \quad \hat{\Omega} = \Omega \tilde{t} \quad (23)$$

where

$$\tilde{t} = \sqrt{\dfrac{(\rho A)_{eq} l^4}{(EI)_{eq}}} \quad (24)$$

Substituting equation (23) in equation (21), and considering viscous damping effect due to the squeeze film damping (Younis et al., 2007) and dropping the hats, the non-dimensional differential equation of the motion reduces to

$$\dfrac{\partial^4 w(x,t)}{\partial x^4} + \dfrac{\partial^2 w(x,t)}{\partial t^2}$$
$$- [\alpha_1 + \alpha_2 \Gamma(w,w)] \dfrac{\partial^2 w(x,t)}{\partial x^2} + \alpha_3 \dfrac{\partial w(x,t)}{\partial t} \quad (25)$$
$$= \alpha_4 \left( \dfrac{[V_{DC} + V_{AC} \sin(\Omega t)]^2}{(1-w)^2} - \dfrac{V_{DC}^2}{(1+w)^2} \right)$$

Where the following non-dimensional boundary conditions holds:

$$w(0,t) = w(1,t) = 0, \quad \dfrac{\partial w(0,t)}{\partial x} = \dfrac{\partial w(1,t)}{\partial x} = 0 \quad (26)$$

For simplicity in equation (25) the function $\Gamma$ and the coefficients $\alpha_i$ are defined as

$$\Gamma(f_1(x,t), f_2(x,t)) = \int_0^1 \dfrac{\partial f_1}{\partial x} \dfrac{\partial f_2}{\partial x} dx$$
$$\alpha_1 = \dfrac{F_P l^2}{(EI)_{eq}}, \quad \alpha_2 = \dfrac{(EA)_{eq} g_0^2}{2(EI)_{eq}}, \quad (27)$$
$$\alpha_3 = \dfrac{\hat{c} \, l^2}{\sqrt{(\rho A)_{eq}(EI)_{eq}}}, \quad \alpha_4 = \dfrac{\varepsilon_0 a l^4}{2 g_0^3 (EI)_{eq}}$$

## 3. Numerical solution

The strong nonlinearity due to the electro-static force and stretching term in equation (25) has often been reported as a challenging dilemma in the literature (Younis et al., 2003). Both sides of equation (25) are multiplied in the denominator of the electrostatic force, which yields

$$(1-w)^2(1+w)^2\frac{\partial^4 w}{\partial x^4} + (1-w)^2(1+w)^2\frac{\partial^2 w}{\partial t^2}$$
$$- (1-w)^2(1+w)^2[\alpha_1 + \alpha_2\Gamma(w,w)]\frac{\partial^2 w}{\partial x^2}$$
$$+ (1-w)^2(1+w)^2\alpha_3\frac{\partial w}{\partial t}$$
$$= \alpha_4 V_{DC}^2[(1+w)^2-(1-w)^2]$$
$$+ (1+w)^2 2\alpha_4 V_{DC} V_{AC}\sin(\Omega t)$$
$$+ \alpha_4(1+w)^2 V_{AC}^2\sin^2(\Omega t) \quad (28)$$

Equation (28) reduces to

$$(1-2w^2+w^4)\frac{\partial^4 w}{\partial x^4} + (1-2w^2+w^4)\frac{\partial^2 w}{\partial t^2}$$
$$- (1-2w^2+w^4)[\alpha_1+\alpha_2\Gamma(w,w)]\frac{\partial^2 w}{\partial x^2}$$
$$+ (1-2w^2+w^4)\alpha_3\frac{\partial w}{\partial t}$$
$$= \alpha_4 V_{DC}^2(4w) + (1+2w+w^2)$$
$$\times (2\alpha_4 V_{DC} V_{AC}\sin(\Omega t) + \alpha_4 V_{AC}^2\sin^2(\Omega t)) \quad (29)$$

Based on the Galerkin method the following approximate solution is substituted in equation (29):

$$w(x,t) = \sum_{m=1}^M q_m(t)\varphi_m(x) \quad (30)$$

where $q_m(t)$ is the $m$th generalized coordinate and $\varphi_m(x)$ is the $m$th normalized linear un-damped mode shape of the straight micro-beam.

Applying Galerkin technique and considering $\varphi_m^{IV}(x) = \alpha_1\varphi_m'' + \omega_m^2\varphi_m$, where $\omega_m$ is the $m$th natural frequency of the straight micro-beam (Appendix A), and based on the orthogonality of the normalized mode shapes as $\int_0^1 \varphi_m(x)\varphi_n(x)dx = \delta_{ij}$, the reduced order differential equation of the micro-beam reduces to

$$\omega_n^2 q_n + \alpha_1\sum_{m=1}^M q_m \int_0^1 \varphi_n\varphi_m''dx - 2\alpha_1\sum_{i=1}^M\sum_{j=1}^M\sum_{m=1}^M q_iq_jq_m\int_0^1 \varphi_n\varphi_i\varphi_j\varphi_m''dx - 2\sum_{i=1}^M\sum_{j=1}^M\sum_{m=1}^M \omega_m^2 q_iq_jq_m$$
$$\times \int_0^1 \varphi_n\varphi_i\varphi_j\varphi_m dx + \alpha_1\sum_{i=1}^M\sum_{j=1}^M\sum_{k=1}^M\sum_{l=1}^M\sum_{m=1}^M q_iq_jq_kq_lq_m\int_0^1 \varphi_n\varphi_i\varphi_j\varphi_k\varphi_l\varphi_m''dx$$
$$+ \sum_{i=1}^M\sum_{j=1}^M\sum_{k=1}^M\sum_{l=1}^M\sum_{m=1}^M \omega_m^2 q_iq_jq_kq_lq_m\int_0^1 \varphi_n\varphi_i\varphi_j\varphi_k\varphi_l\varphi_m dx + \ddot{q}_n - 2\sum_{i=1}^M\sum_{j=1}^M\sum_{m=1}^M q_iq_j\ddot{q}_m\int_0^1 \varphi_n\varphi_i\varphi_j\varphi_m dx$$
$$+ \sum_{i=1}^M\sum_{j=1}^M\sum_{k=1}^M\sum_{l=1}^M\sum_{m=1}^M q_iq_jq_kq_l\ddot{q}_m\int_0^1 \varphi_n\varphi_i\varphi_j\varphi_k\varphi_l\varphi_m dx - \alpha_1\sum_{m=1}^M q_m\int_0^1 \varphi_n\varphi_m''dx + 2\alpha_1\sum_{i=1}^M\sum_{j=1}^M\sum_{m=1}^M q_iq_jq_m\int_0^1 \varphi_n\varphi_i\varphi_j\varphi_m''dx$$
$$- \alpha_1\sum_{i=1}^M\sum_{j=1}^M\sum_{k=1}^M\sum_{l=1}^M\sum_{m=1}^M q_iq_jq_kq_lq_m\int_0^1 \varphi_n\varphi_i\varphi_j\varphi_k\varphi_l\varphi_m''dx - \alpha_2\sum_{m=1}^M\sum_{p=1}^M\sum_{r=1}^M q_mq_pq_r\int_0^1 \varphi_n\varphi_m''\int_0^1 \varphi_p'\varphi_r'dxdx$$
$$+ 2\alpha_2\sum_{i=1}^M\sum_{j=1}^M\sum_{m=1}^M\sum_{p=1}^M\sum_{r=1}^M q_iq_jq_mq_pq_r\int_0^1 \varphi_n\varphi_m''\varphi_i\varphi_j\int_0^1 \varphi_p'\varphi_r'dxdx - \alpha_2\sum_{i=1}^M\sum_{j=1}^M\sum_{k=1}^M\sum_{l=1}^M\sum_{m=1}^M\sum_{p=1}^M\sum_{r=1}^M q_iq_jq_kq_lq_mq_pq_r$$
$$\times \int_0^1 \varphi_n\varphi_i\varphi_j\varphi_k\varphi_l\varphi_m''\int_0^1 \varphi_p'\varphi_r'dxdx + \alpha_3\dot{q}_n - 2\alpha_3\sum_{i=1}^M\sum_{j=1}^M\sum_{m=1}^M q_iq_j\dot{q}_m\int_0^1 \varphi_n\varphi_i\varphi_j\varphi_m dx$$
$$+ \alpha_3\sum_{i=1}^M\sum_{j=1}^M\sum_{k=1}^M\sum_{l=1}^M\sum_{m=1}^M q_iq_jq_kq_l\dot{q}_m\int_0^1 \varphi_n\varphi_i\varphi_j\varphi_k\varphi_l\varphi_m dx - 2\alpha_4[(V_{DC}+V_{AC}\sin(\Omega t))^2+V_{DC}^2]q_n$$
$$- \alpha_4[(V_{DC}+V_{AC}\sin(\Omega t))^2-V_{DC}^2]\times\sum_{i=1}^M\sum_{j=1}^M q_iq_j\int_0^1 \varphi_n\varphi_i\varphi_j dx = \alpha_4[(V_{DC}+V_{AC}\sin(\Omega t))^2-V_{DC}^2]\int_0^1 \varphi_n dx$$

$$(31)$$

To find a period solution of equation (31), we have applied the shooting method (Nayfeh and Balachandran, 1995; Nayfeh et al., 2007b). Once a periodic solution is captured, its stability is investigated by examining the eigenvalues of the monodromy matrix (Nayfeh and Balachandran, 1995; Younis, 2010). The monodromy matrix is evaluated as a by-product of the shooting method.

## 4. Results and discussions

As a case study we have studied a fully clamped FGP micro beam (as indicated in Figure 1) with the mechanical and geometrical properties listed in Table 1 (Rezazadeh et al., 2009).

Figure 2 (a–c) depicts the distribution of the mechanical/electrical properties including elasticity modulus, density, and piezoelectric voltage constant along the thickness of the FGP micro-beam. It is worth mentioning that the mid-plane of the micro-beam is of pure silicon ($P_{p_0} = 0.00$) while the properties are assumed to be symmetrically distributed according to the power law distribution (Azizi et al., 2012) on either sides of the mid-plane.

Associated with $P_{p_u} = 0.00$ the micro-beam becomes pure silicon and as a result the mechanical/electrical properties remain constant along the height ($z$ axis).

In dynamical systems with either external or parametric excitation, identification of the periodic solutions and investigation of their stability is critically important. The concept of fixed points in dynamical systems with homogenous differential equations can be generalized to the concept of periodic solutions in systems with either external or parametric excitations because both fixed points and periodic solutions show the possible responses/attractors as the transient response vanishes.

Associated with the system of equations (31), based on the initial conditions, the system either undergoes a bounded periodic response or an ever-increasing response which results in pull-in. From the point of view that the frequency response curves illustrate the steady-state responses, determination of frequency response curves is essential. A periodic solution of a stable type on the frequency response curve plays the role of an attractor and can be a possible steady-state response.

**Table 1.** Geometrical and material properties of the micro-beam and piezoelectric layers.

| Geometrical and material properties | Silicon | Piezoelectric material (PZT) |
|---|---|---|
| Length | 600 μm | |
| Width | 30 μm | |
| Thickness | 1 μm | |
| Initial gap | 2 μm | |
| Young's modulus | 169.61 GPa | 76.6 GPa |
| Density | 2331 kg/m³ | 7500 kg/m³ |
| $e_{31}$ | – | – 9.29 Coulomb/m² |

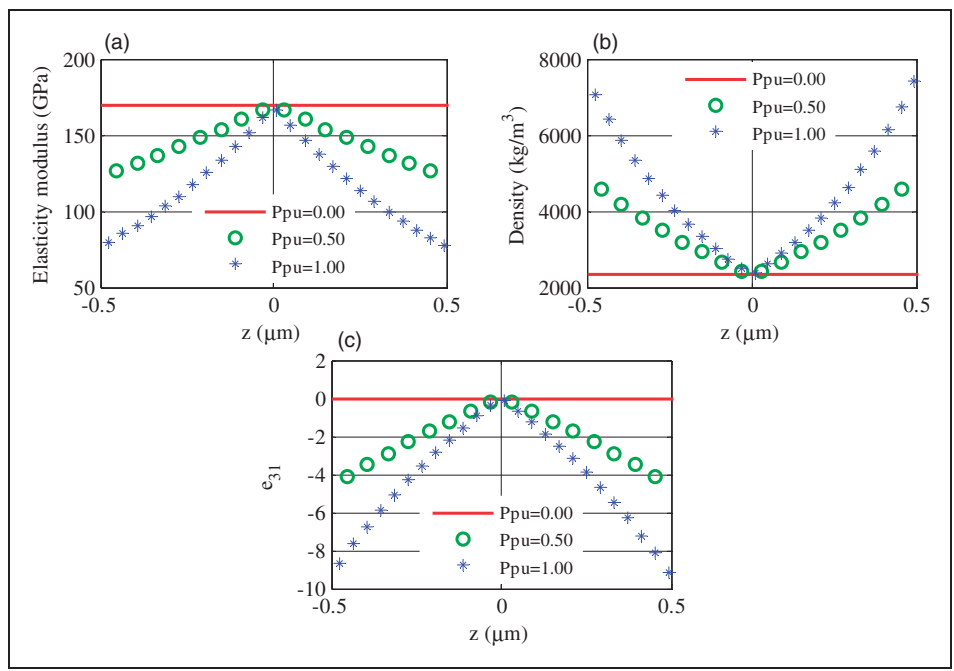

**Figure 2.** Distribution of the mechanical properties along the thickness of the FGP micro-beam with $P_{p_0} = 0.00$ and various $P_{p_u}$, (a) elasticity modulus, (b) density, (c) piezoelectric voltage constant.

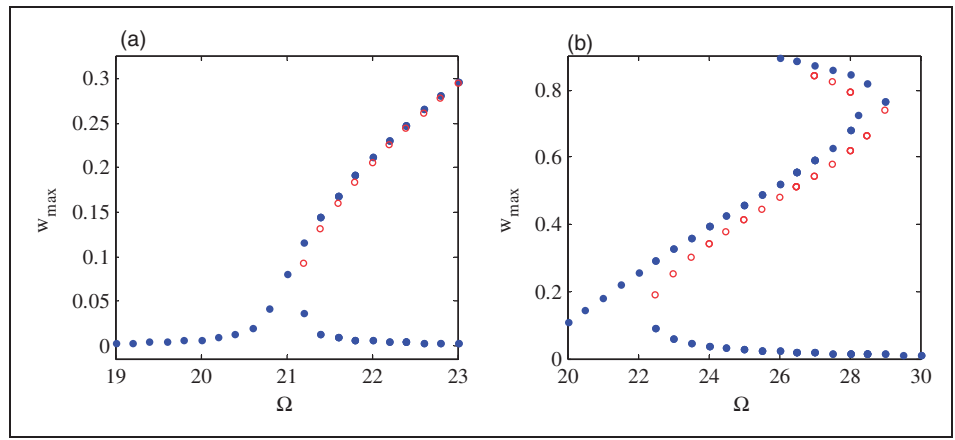

**Figure 3.** Frequency response curve in the vicinity of primary resonance (filled circles represent the stable periodic solutions), with $V_{DC} = 2.0$ V, $P_{p_u} = 0.00$, (a): $V_{AC} = 10$ mV, (b): $V_{AC} = 200$ mV.

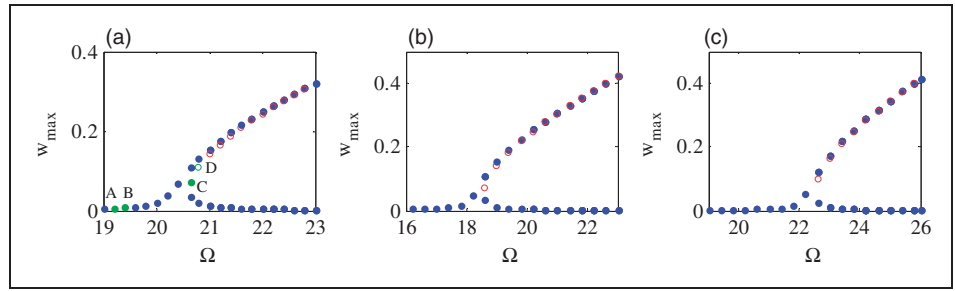

**Figure 4.** Frequency response curve representing the hardening effect near primary resonance (filled circles represent the stable periodic solutions) $V_{DC} = 2.0$ V, $V_{AC} = 10$ mV, $P_{p_u} = 0.50$; (a) $V_P = 0.0$ mV, (b) $V_P = -90.0$ mV, (c) $V_P = 90.0$ mV.

The frequency response curves corresponding to various piezoelectric and electrostatic voltages and different $P_{p_u}$s have been illustrated in following figures.

Figure 3 depicts the frequency response curves corresponding to $P_{p_u} = 0.00$, which implies pure silicone micro-beam.

The frequency response curves show that for low amplitudes of harmonic excitation in the vicinity of the primary resonance the system exhibits hardening behavior as the frequency is swept forward. This behavior is due to the hardening effect of the geometric nonlinearity (mid-plane stretching) which resembles a nonlinear cubic type of stiffness (Younis, 2010). As mentioned, two main sources of nonlinearity including geometric nonlinearity and nonlinear electro-static force govern the dynamics of the micro-beam.

The geometric nonlinearity has a hardening effect while the electrostatic force has a softening effect (Nayfeh et al., 2007a; Younis et al., 2003). As the amplitude of the harmonic excitation increases, the amplitude of the periodic solutions also increases; this leads to the domination of the softening effect of the electrostatic force on the dynamics of the micro-resonator and accordingly the appearance of asoftening nature in the frequency response curve.

Figure 4 represents the frequency response curves for the case of $P_{p_u} = 0.50$ and $V_{DC} = 2$ v $V_{AC} = 10$ mv and various piezoelectric actuations.

As the frequency response curves depict, based on the polarity of piezoelectric actuation, the frequency response curves can be tuned both in forward and backward directions. Application of piezoelectric voltage with positive polarity leads to the generation of tensional axial force and as a result the frequency response curves move forward along the frequency axis in compared with $V_P = 0.0$ mv. Piezoelectric actuation with negative polarity imposes a compressive axial force along the micro-beam and accordingly the frequency response curve moves backward in compared with $V_P = 0.0$ mv. Figure 5 illustrates the frequency response curves for the same $V_{DC}$, and $P_{p_u}$ as Figure 4, and $V_{AC} = 200$ mv, and various piezoelectric voltages.

Comparing the frequency response curves depicted in Figures 4 and 5 reveal that both softening and hardening effects appear for higher $V_{AC}$s whereas for lower amplitudes of harmonic excitation the micro-resonator exhibits a hardening effect – this qualitative behavior is also reported in the literature (Nayfeh et al., 2007a; Azizi et al., 2013). Furthermore for lower amplitudes of harmonic excitation there is one cyclic fold bifurcation, but higher amplitudes lead to three cyclic fold bifurcation points on the frequency response curves.

Figure 6 illustrates the Floquet exponents appertaining to the periodic solutions of Figure 4(a) as the control parameter (excitation frequency) varies from $A$ to $D$. The Floquet exponents corresponding to point $A$ are all located inside the unit circle, meaning that the periodic attractor is stable. As we move forward on the frequency response curve the imaginary parts of the Floquet exponents become smaller and they move toward the real axis on the complex plane. At point $C$, a cyclic fold type of bifurcation occurs and the system exhibits multi-response. This type of bifurcation is of discontinuous or catastrophic type and the corresponding Floquet exponents approach unity meaning that the periodic solution is of non-hyperbolic type and the linearization breaks down to predict the stability. As we move further on the frequency response curve (point *D*) three steady-state solutions including two stable and one unstable appear. One of the Floquet exponents associated with the unstable solution is inside and the other one is outside the unit circle. In other words the periodic solution is of saddle type having one stable and one unstable manifold and any perturbations result in the pull-in phenomenon unless the applied perturbation is on the stable manifold; however this is only of mathematical interest rather than physical interest. Pull-in is a snap-through like behavior and is a saddle-node bifurcation type of instability (Zhang and Zhao, 2006).

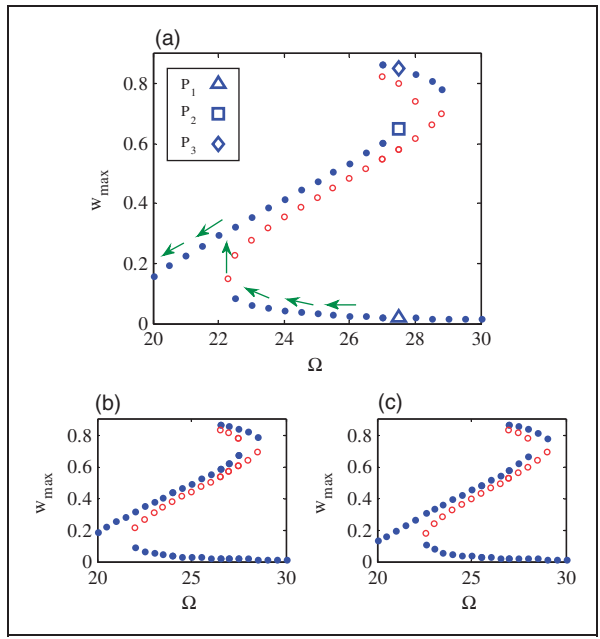

**Figure 5.** Frequency response curve representing the hardening effect near primary resonance (filled circles represent the stable periodic solutions) $V_{DC} = 2.0$ V, $V_{AC} = 200$ mV, $P_{p_u} = 0.50$; (a) $V_P = 0.0$ mV, (b) $V_P = -20.0$ mV, (c) $V_P = 20.0$ mV.

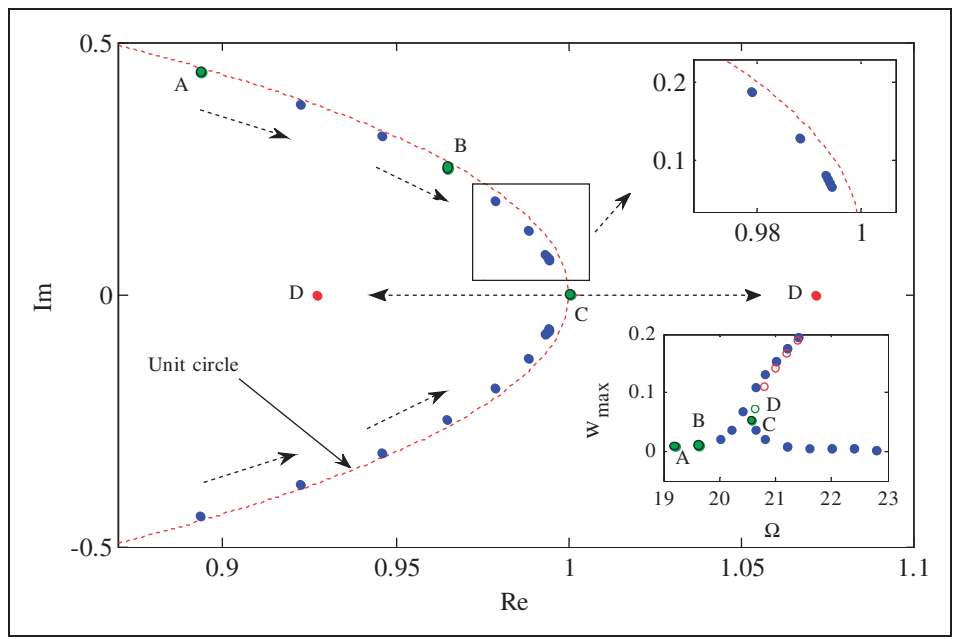

**Figure 6.** The Floquet exponents and the unit circle (dashed) on the complex plane associated with the parameters of Figure 4(a). (The frequency response curve is shown as inset).

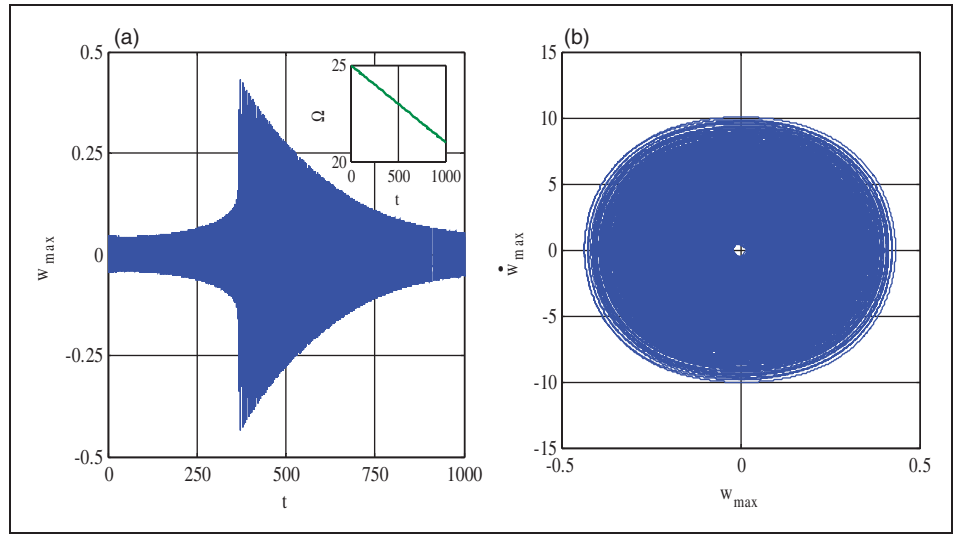

**Figure 7.** (a) Temporal response, (b) the corresponding phase plane associated with the mid-point of the micro-beam subjected to backward frequency sweep, $V_{DC} = 2.0$ V, $V_{AC} = 200$ mV, $V_P = 0.0$ mV.

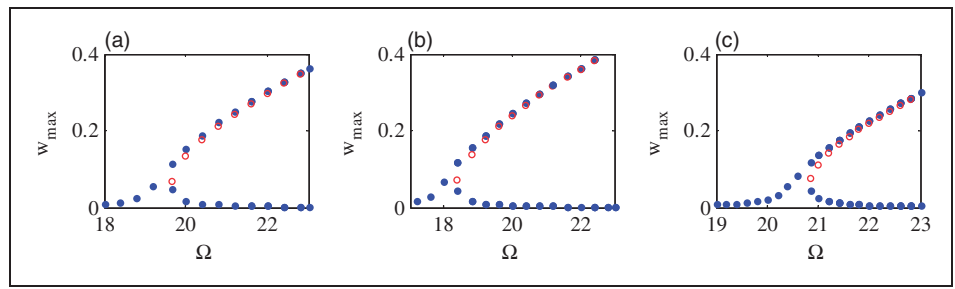

**Figure 8.** Frequency response curve representing the hardening effect near primary resonance (filled circles represent the stable periodic solutions) $V_{DC} = 2$ V, $V_{AC} = 10$ mV, $P_{p_u} = 1.00$; (a) $V_P = 0.0$ mV, (b) $V_P = -20.0$ mV, (c) $V_P = 20.0$ mV.

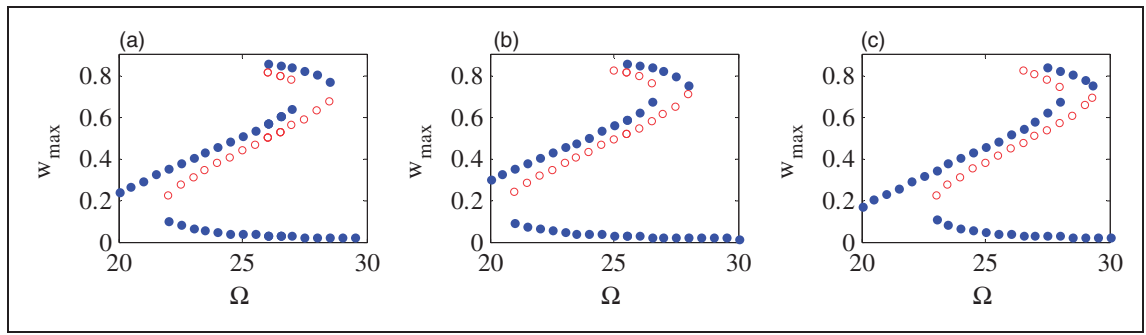

**Figure 9.** Frequency response curve representing the hardening effect near primary resonance (filled circles represent the stable periodic solutions) $V_{DC} = 2.0$ V, $V_{AC} = 200$ mV, $P_{p_u} = 1.00$; (a) $V_P = 0.0$ mV, (b) $V_P = -20.0$ mV, (c) $V_P = 20.0$ mV.

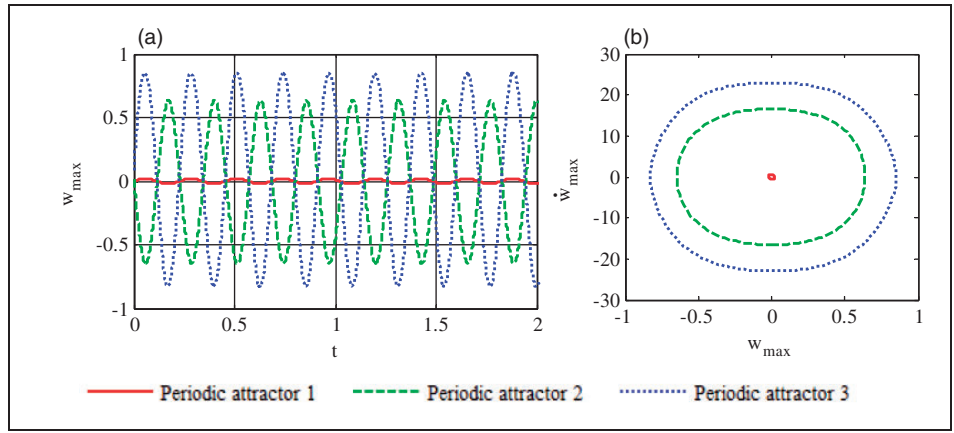

**Figure 10.** Periodic attractors corresponding to $V_{DC} = 2.0$ V, $V_{AC} = 200$ mV, $V_P = 0.0$ mV, $P_{p_u} = 0.50$, $\Omega = 27.5$ (a) Temporal response, (b) phase plane.

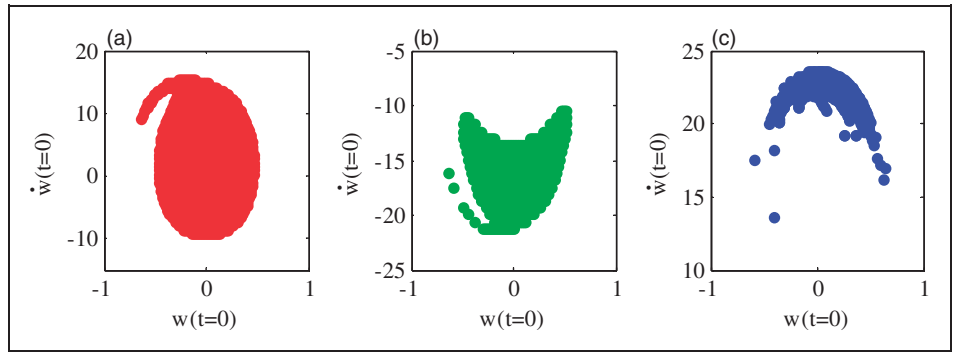

**Figure 11.** The basins of the periodic attractors corresponding to $V_{DC} = 2$ V, $V_{AC} = 200$ mV, $V_P = 0.0$ mV, $P_{p_u} = 0.50$, $\Omega = 27.5$ (a) $P_1$, (b) $P_2$, (c) $P_3$.

Figure 7 represents the temporal response and the phase plane of the midpoint of the micro-beam subjected to backward frequency sweep with $V_{DC} = 2v$, $V_{AC} = 200mv$, $V_P = 0.0mv$ (the appertaining frequency response curve is depicted in Figure 5(a). As the frequency is swept from $\Omega = 25$ to $\Omega = 21$, the response is attracted to the lower amplitude limit cycle appertaining to $\Omega = 25$ and decreasing the frequency, results in the increase of the amplitude of the motion which is in agreement with what the frequency response curves predict (Figure 5(a)). At $\Omega = 22.3$ the response is suddenly attracted to the stable periodic solution with higher amplitude, and further reduction of the excitation frequency leads to the reduction of the amplitude of the motion.

Figure 8 represents the frequency response curve for the case of $V_{DC} = 2$ v $V_{AC} = 10$ mv, $P_{p_u} = 1.00$ and various applied piezoelectric voltages.

Figure 9 represents the frequency response curves for the same values of the parameters as those of Figure 8 but with $V_{AC} = 200$ mv.

Comparing the frequency response curves of Figures 5 and 9 reveals that the frequency shift for $P_{p_u} = 1.00$ in comparison with the frequency shift for the case where $P_{p_u} = 0.50$ is considerably more; for a definite piezoelectric voltage the more the $P_{p_u}$, the more is the axial force and accordingly the more is the frequency shift. Tunability of the frequency response curves enables the design of tunable resonant sensors or RF filters.

Figure 10, depicts three periodic attractors illustrated in Figure 5(a). The initial conditions are exactly chosen on the periodic attractors so that the resonator does not undergo transient responses.

In nonlinear systems corresponding to a determined excitation frequency, the system may have multi attractors (known as stable periodic solutions or stable limit cycles), assuming a bounded response condition, the steady-state response, and the amplitude of the steady-state response depends on the fact that the applied initial condition is located on the basin of which attractor. Figure 11, depicts the basins of attractions for the three stable periodic solutions (attractors) corresponding to $\Omega = 27.5$ which are illustrated in Figure 5(b).

The basin of attraction becomes smaller as the amplitude of the periodic attractor increases; this behavior is in agreement with the experimental analysis of Younis et al. (2010) and theoretical investigation of Azizi et al. (2014).

## 5. Conclusion

The study focused on the nonlinear dynamics of an FGP micro-resonator subject to a two-sided electrostatic and a simultaneous piezoelectric actuation as a micro-resonator. The shooting method was applied to detect the periodic. Two main sources of nonlinearities including electrostatic force and geometric nonlinearity due to mid-plane stretching were governing the global behavior of the proposed micro-resonator. The dynamics of the micro-resonator in the vicinity of the primary resonance were studied; for low enough amplitudes of the harmonic excitation, the effect of geometric nonlinearity was dominant and as a result the frequency response curves were of the hardening type; regardless of the polarity of the piezoelectric voltage we saw one cyclic-fold bifurcation in the frequency response curves where the Floquet exponents laid on the unit circle and the periodic orbits became of a non-hyperbolic type. Increasing the amplitude of the harmonic excitation led to the increase of the amplitude of the periodic solutions, and as a result the softening effect of electrostatic force dominated the hardening effect of geometric nonlinearity. The micro-resonator underwent three cyclic-fold bifurcations for high amplitudes of harmonic excitations. Applying piezoelectric voltage resulted in a compressive/tensional axial force along the length of the micro-resonator; this made the operating frequency range of the micro-resonator tunable. Based on the polarity of the piezoelectric voltage we could shift the operating frequency in both forward and backward directions. The Floquet exponents corresponding to stable periodic orbits were all located inside the unit circle, however at least one of the Floquet exponents of the saddle type unstable periodic orbits were outside the unit circle. We explored the basins of attraction corresponding to three possible attractors for a definite actuation frequency; the area of the attractor on the phase space was inversely proportional to the amplitude of the attractor. The results of the present study can be used as a design tool for tunable micro-resonators.

## Funding

This research received no specific grant from any funding agency in the public, commercial, or not-for-profit sectors.

# Appendix A

$$\alpha_1 \sum_{m=1}^{M} q_m \int_0^1 \varphi_n \varphi_m'' dx + \sum_{m=1}^{M} q_m \omega_m^2 \int_0^1 \varphi_n \varphi_m dx - 2\alpha_1 \sum_{i=1}^{M} \sum_{j=1}^{M} \sum_{m=1}^{M} q_i q_j q_m \int_0^1 \varphi_n \varphi_i \varphi_j \varphi_m'' dx$$

$$- 2 \sum_{i=1}^{M} \sum_{j=1}^{M} \sum_{m=1}^{M} \omega_m^2 q_i q_j q_m \int_0^1 \varphi_n \varphi_i \varphi_j \varphi_m dx + \alpha_1 \sum_{i=1}^{M} \sum_{j=1}^{M} \sum_{k=1}^{M} \sum_{l=1}^{M} \sum_{m=1}^{M} q_i q_j q_k q_l q_m \int_0^1 \varphi_n \varphi_i \varphi_j \varphi_k \varphi_l \varphi_m'' dx$$

$$+ \sum_{i=1}^{M} \sum_{j=1}^{M} \sum_{k=1}^{M} \sum_{l=1}^{M} \sum_{m=1}^{M} \omega_m^2 q_i q_j q_k q_l q_m \int_0^1 \varphi_n \varphi_i \varphi_j \varphi_k \varphi_l \varphi_m dx + \sum_{m=1}^{M} \ddot{q}_m \int_0^1 \varphi_n \varphi_m dx - 2 \sum_{i=1}^{M} \sum_{j=1}^{M} \sum_{m=1}^{M} q_i q_j \ddot{q}_m \int_0^1 \varphi_n \varphi_i \varphi_j \varphi_m dx$$

$$+ \sum_{i=1}^{M} \sum_{j=1}^{M} \sum_{k=1}^{M} \sum_{l=1}^{M} \sum_{m=1}^{M} q_i q_j q_k q_l \ddot{q}_m \int_0^1 \varphi_n \varphi_i \varphi_j \varphi_k \varphi_l \varphi_m dx - \alpha_1 \sum_{m=1}^{M} q_m \int_0^1 \varphi_n \varphi_m'' dx + 2\alpha_1 \sum_{i=1}^{M} \sum_{j=1}^{M} \sum_{m=1}^{M} q_i q_j q_m \int_0^1 \varphi_n \varphi_i \varphi_j \varphi_m'' dx$$

$$- \alpha_1 \sum_{i=1}^{M} \sum_{j=1}^{M} \sum_{k=1}^{M} \sum_{l=1}^{M} \sum_{m=1}^{M} q_i q_j q_k q_l q_m \int_0^1 \varphi_n \varphi_i \varphi_j \varphi_k \varphi_l \varphi_m'' dx - \alpha_2 \sum_{m=1}^{M} \sum_{p=1}^{M} \sum_{r=1}^{M} q_m q_p q_r \int_0^1 \varphi_n \varphi_m'' \Gamma(\varphi_p, \varphi_r) dx$$

$$+ 2\alpha_2 \sum_{i=1}^{M} \sum_{j=1}^{M} \sum_{m=1}^{M} \sum_{p=1}^{M} \sum_{r=1}^{M} q_i q_j q_m q_p q_r \int_0^1 \varphi_n \varphi_m'' \varphi_i \varphi_j \Gamma(\varphi_p, \varphi_r) dx - \alpha_2 \sum_{i=1}^{M} \sum_{j=1}^{M} \sum_{k=1}^{M} \sum_{l=1}^{M} \sum_{m=1}^{M} \sum_{p=1}^{M} \sum_{r=1}^{M} q_i q_j q_k q_l q_m q_p q_r$$

$$\times \int_0^1 \varphi_n \varphi_i \varphi_j \varphi_k \varphi_l \varphi_m'' \Gamma(\varphi_p, \varphi_r) dx + \alpha_3 \sum_{m=1}^{M} \dot{q}_m \int_0^1 \varphi_n \varphi_m dx - 2\alpha_3 \sum_{i=1}^{M} \sum_{j=1}^{M} \sum_{m=1}^{M} q_i q_j \dot{q}_m \int_0^1 \varphi_n \varphi_i \varphi_j \varphi_m dx$$

$$+ \alpha_3 \sum_{i=1}^{M} \sum_{j=1}^{M} \sum_{k=1}^{M} \sum_{l=1}^{M} \sum_{m=1}^{M} q_i q_j q_k q_l \dot{q}_m \int_0^1 \varphi_n \varphi_i \varphi_j \varphi_k \varphi_l \varphi_m dx - 2\alpha_4 (V_{DC} + V_{AC} \sin(\Omega t))^2 \sum_{i=1}^{M} q_i \int_0^1 \varphi_n \varphi_i dx$$

$$- \alpha_4 (V_{DC} + V_{AC} \sin(\Omega t))^2 \sum_{i=1}^{M} \sum_{j=1}^{M} q_i q_j \int_0^1 \varphi_n \varphi_i \varphi_j dx - 2\alpha_4 V_{DC}^2 \sum_{i=1}^{M} q_i \int_0^1 \varphi_n \varphi_i dx + \alpha_4 V_{DC}^2$$

$$\times \sum_{i=1}^{M} \sum_{j=1}^{M} q_i q_j \int_0^1 \varphi_n \varphi_i \varphi_j dx = \alpha_4 (V_{DC} + V_{AC} \sin(\Omega t))^2 \int_0^1 \varphi_n dx - \alpha_4 V_{DC}^2 \int_0^1 \varphi_n dx$$

$$(A-1)$$